\documentclass[11pt]{iopart}
\usepackage{epstopdf}
\usepackage[dvipdfm]{graphicx}
\usepackage{tabularx}
\usepackage{color}
\usepackage{amssymb,amsthm}
 \usepackage{footnote}
\def\eV2{eV$^{2}$}
\def\sinsqtt13{sin$^{2}(2\theta _{13})$}
\def\sin22t{sin$^{2}(2\theta)$}
\def\sin22tnew{sin$^{2}(2\theta _{new})$}

\renewcommand\footnotemark{}
\renewcommand\footnotetext{}

\begin{document}

\title[Study of Neutrino Oscillations in the OPERA experiment]{Study of Neutrino Oscillations in the OPERA experiment
\footnote{Prensented at the Lake Louise Winter 2013 Conference, Banff, Alberta, Canada, 17-23 February 2013.}}
\author{UMUT KOSE\\On behalf of OPERA Collaboration}
\address{ 
  INFN Sezione di Padova,
  \\
  I-35131 Padova, Italy\\
}
 
\eads{\mailto{umut.kose@cern.ch}}



\begin{abstract}
The OPERA experiment has been designed to perform the first detection of 
neutrino oscillations in direct appearance mode in the $\nu_{\mu}\to\nu_{\tau}$ channel, 
through the detection of the tau lepton produced in charged current interaction on an event by event basis. 
The detector is hybrid, being made of an emulsion/lead target and of electronic detectors. 
It exploited the CNGS muon neutrino beam from CERN to Gran Sasso, 730 km from the source.
Runs with CNGS neutrinos were successfully carried out from 2008 to 2012. 
We report on the large data sample analysed so far and give our results on the search for $\nu_{\mu}\to\nu_{\tau}$ and 
$\nu_{\mu}\to\nu_{e}$ oscillations.
\end{abstract}

\normalsize\baselineskip=15pt

\section{Introduction}

Most of existing data on neutrino oscillations from the solar, atmospheric, reactor and accelerator experiments
have established a framework of neutrino oscillations among three flavor neutrinos through the mixing of three mass eigenstates.

Atmospheric sector flavor conversion
was first established by the Super-Kamiokande~\cite{ref:sk} 
and MACRO~\cite{ref:macro} experiments and then confirmed 
by the K2K~\cite{ref:k2k} and MINOS~\cite{ref:minos} long-baseline
experiments. The hypothesis of $\nu_{\mu}\to\nu_{\tau}$ oscillation origin is fitting well all the data.
Moreover, the CHOOZ~\cite{ref:chooz} and Palo Verde~\cite{ref:paloverde} reactor
experiments ruled out indirectly the $\nu_{\mu}\rightarrow\nu_{e}$ channel as the
dominant mode in the atmospheric sector. 
It is important to note that all of the experiments yielded positive evidence
for neutrino oscillations on the disappearance of muon neutrinos from the atmospheric 
neutrino flux, with no explicit observation of the appearance of tau neutrinos and their Charged Current (CC) interactions.

The main goal of the OPERA experiment~\cite{ref:opera}
is to perform a unique appearance observation of the oscillation products to confirm
unambiguously the neutrino oscillation hypothesis in the atmospheric sector 
through the $\nu_{\mu}\to\nu_{\tau}$ channel. 
Although the neutrino beam is not optimized for it, OPERA is performing a study on subleading 
$\nu_{\mu}\to\nu_e$ oscillation search at atmospheric and high $\Delta m^2$.

In the following, the neutrino beam, the OPERA detector with its performances and the event-by-event analysis stream 
will be briefly reviewed. The neutrino oscillation results both on $\nu_{\mu}\to\nu_{\tau}$ and $\nu_{\mu}\to\nu_e$
will be discussed.

\section{The neutrino beam and the OPERA detector}

In order to study $\nu_{\mu}\rightarrow{\nu}_{\tau}$ oscillations in appearance mode 
in the atmospheric neutrino sector, 
the CERN Neutrinos to GranSasso (CNGS) neutrino beam~\cite{ref:cngs} 
was designed and optimized by maximizing the number of $\nu_{\tau}$ CC interactions at the 
Gran Sasso underground laboratory (LNGS)~\cite{ref:lngs}.
The average $\nu_{\mu}$ beam energy
is 17 GeV, well above tau production energy treshold. The $\bar{\nu}_{\mu}$ contamination is $\sim 4\%$ in flux, $2.1\%$ in terms of interactions. The 
$\nu_{e}$ and $\bar{\nu}_{e}$ contaminations are lower than $1\%$, while the number of prompt $\nu_{\tau}$ from
$D_{s}$ decay is negligible, ${\mathcal{O}}(10^{-6})$. 
The average $L/E_{\nu}$ ratio is 43 km/GeV, suitable for oscillation studies at atmospheric
$\Delta{m^2}$. 

The CNGS completed its operation before the 2013 CERN accelerator complex shut-down on December 3, 2012. 
A total number of $17.97\times10^{19}$ protons on target (pots) was
accumulated since the beginning of the program in 2008. 
It corresponds to 80\% of the CNGS beam originally foreseen, $22.5\times10^{19}$ pots in 5 years.

The neutrino beam produced by the CERN-SPS is directed towards the
OPERA detector, located in the LNGS in Italy,
730 km away from the neutrino source. 
The challenge 
of the experiment is the detection of the short lived $\tau^{-}$ lepton (c$\tau$=87 $\mu$m, decay length $\sim$1 mm at the average CNGS beam energy) produced in the CC interaction of
$\nu_\tau$. 

The detector should match a large target mass to collect 
enough statistics and a very high spatial
resolution to observe the $\tau$ lepton with high rejection power to limit background contamination.
Therefore, the OPERA detector was designed as an hybrid detector made of two identical 
Super Modules (SM1 and SM2), each
one formed by a target section and a muon spectrometer. The muon spectrometer is used to reconstruct
and identify muons from $\nu_\mu$ CC interactions and measure their momentum and charge.
Each target section is organized
in 31 vertical ``$walls$'', transverse to the beam direction.
Walls are filled with Emulsion Cloud Chamber units,``$ECC$ $bricks$'', with an
overall mass of 1.25 kton. Each wall is followed by
double layers of scintillator planes, the Target
Trackers (TT), used to locate neutrino
interactions occurred within the target.
An ECC brick consists of 56 lead
plates of 1 mm thickness interleaved with 57
emulsion films. 
The lead plates serve as neutrino interaction
target and the emulsion films as three-dimensional (3D) tracking
detectors providing track coordinates with a sub-micron
accuracy and track angles with a few mrad accuracy which is ideal for $\nu_{\tau}$ interaction
detection.
The material of a brick along the beam direction corresponds
to about 10 radiation lengths and 0.33 interaction
length. The brick size is 10$\times$12.5$\times$8 cm$^3$ and its weight is
about 8.3 kg.

The brick as stand alone detector also allows the momentum measurement of charged particles by
Multiple Coulomb Scattering (MCS) and the separation of low energy $\pi$ and $\mu$ by $dE/dx$
measurements. Electromagnetic showers can be detected and unambigouously attributed to
electrons, $\gamma$s or $\pi^{0}$.

The experimental apparatus and techniques are described in detail elsewhere~\cite{ref:opera}.

\section{Event analysis in OPERA}

All the events triggered by the neutrino beam spill the so-called ``on-time'' events, in coincidence with the 
two 10.5~$\mu$s long CNGS spills distant of 50~ms are considered for neutrino oscillation studies.
Neutrino event analysis starts with the pattern recognition in the electronic detectors. 
Charged particle tracks
produced in a neutrino interaction generate signals in the TT and in the muon spectrometer. These signals are used 
to classify the event 
as CC-like or Neutral Current (NC), like interactions based on the presence of at least
a 3D reconstructed track idenfied as a muon in the event\footnote{if 
the reconstructed 3D
track has a length times density value more than 660 g$\times$cm${^{-2}}$, the event is classified as CC-like.}.
 
 An automatic algorithm is applied to all on-time events in order to select the events contained 
in the emulsion target. A brick finding algorithm is applied then to select the brick with the
maximum probability to contain the neutrino interaction which
 is extracted from the detector by an automated system, the
Brick Manipulator System. 

After extraction, the brick is validated by analysing the Changeable Sheet (CS)~\cite{ref:CSpaper} doublet of
films externally attached to it. 
The signal recorded in the CS films confirms the prediction of the electronic detector, or acts as veto and triggers
a search in neighbouring bricks to find the correct ECC where the neutrino interaction has occured.
The measurement of emulsion films is performed
through high-speed automated microscopes~\cite{ref:ess,ref:suts} with a sub-micrometric position
resolution and an angular resolution of the order of
one milliradian. 

If there is not any CS track compatible with the electronic detector tracks related to the event,
 the brick is returned back to
the detector with a new CS doublet attached and the next brick in the probability list is extracted. 
This method allows to save the analysis time of the brick and minimize the loss of target mass.
If any track originating from the interaction is
detected in the CS, the brick is exposed to cosmic
rays (for alignment purposes) and then depacked.
The emulsion films are developed and
sent to the scanning laboratories of the Collaboration
for event location studies and decay search
analysis. 

\subsection{Event Location}

All track information from the CS analysis is used for a precise
prediction of the tracks in the most
downstream film of the brick (with an accuracy of about 100 $\mu$m). 
When found in this film, tracks are followed back
from film to film. The scan-back procedure
is stopped when no track candidate is found in
three consecutive films; the lead plate just upstream
the last detected track segment is defined as the vertex
plate. In order to study the located vertices and reconstruct the events, 
a general scanning volume is defined with a transverse area of 1$\times$1 cm$^{2}$ along 5 films upstream and 10 films
downstream of the stopping point. All track segments in this volume are collected and analysed. After rejection of
the passing through tracks related to cosmic rays and of the tracks due to low energy particles, the tracks produced by the neutrino
interaction can be selected and reconstructed.

\subsection{Decay Search}\label{decaysearch}
Once the neutrino interaction is located (mainly in lead plates), a decay search procedure is applied to detect
possible short/long decays or interaction topologies on tracks attached to the primary vertex.
For short decays, about 60\%, the $\tau$ decays in the same lead plate in which the neutrino interaction occurs.
The main signature of a $\tau$ candidate is the observation of a track with
a significant impact parameter (IP) relative to the neutrino
interaction vertex. The IP of primary tracks is smaller than 10 $\mu$m after excluding tracks 
produced by low momentum particles.
For long decays, about 40\%, the candidate event is 
selected on the basis either of the existence of a clear ``kink''($\theta_{kink}$ \textgreater 20 mrad)
or of a secondary multiprong vertex
from the decay in flight of a particle emerging from the
neutrino interaction vertex.

When secondary vertices or kink topologies
are found in the event, a kinematical analysis
is performed, based on particle angles and momenta
measured in the emulsion films. For charged particles
up to about 6 GeV/c, momenta can be determined
using the angular deviations produced
by MCS of tracks
in the lead plates with a resolution better
than $22\%$~\cite{ref:mcs}. 
For higher momentum particles, the measurement is based
on the position deviations. The resolution is better
than $33\%$ on 1/p up to 12 GeV/c for particles crossing
 an entire brick.

A $\gamma$-ray search is performed in
the whole scanned volume by checking all tracks
having an IP with respect to
the primary or secondary vertices smaller than 800 $\mu$m. The angular acceptance is $\pm500$ mrad. The
$\gamma$-ray energy is estimated by a Neural Network algorithm
that uses the number
of segments, the shape of the electromagnetic
shower and also the MCS of the leading particles.

The identification of an electron is essentially based on the detection of 
the associated electromagnetic shower. Since the size of the standard scanning volume is generally too short
in the beam direction to contain the electromagnetic shower, a special electron-ID scan is performed on NC-like located events
by extending the scanning volume. All primary tracks are extrapolated to the CS and the existence of electromagnetic shower
associated to that tracks are investigated. If 3 or more tracks are found in the CS within 150 mrad angular and 2 mm spatial
tolerances, around a given primary track, an additional volume along the candidate track is scanned.
If the automatic algorithm based on a Neural Network reconstructs an electromagnetic shower, and if the primary track 
initiating the shower is recognized as a single track on the vertex emulsion film, the event is classified as a $\nu_e$ candidate event.

\section{Neutrino Oscillations results}

The results are based on the analysis
of the first two years of data (2008-2009), which is completed~\cite{ref:20082009} , and of a 2010-2011
data sub-sample. For the sample currently under analysis and accumulated during the runs from
2010 to 2012, a pre-selection is applied that includes all NC-like events and CC-like events with a muon momentum
below 15 GeV/c.
At the time of the conference, 4505 neutrino interactions were fully analysed looking for decay topologies: two $\nu_\tau$
neutrino events were found. At the time of writing these proceedings, with the extension of the
analysed sample to events with one $\mu$ in the final state a third $\tau$ candidate, 
coming from $\tau$ lepton decays to muon channel, was detected~\cite{ref:thirdtau}. 

A systematic search of $\nu_e$ interaction was applied to 505 NC-like located events from 2008-2009 data, 
19 $\nu_e$ candidates were detected.

In the following section, the results of the oscillation analyses are presented.

\subsection{$\nu_{\mu}\to\nu_{\tau}$ oscillations}

The $\nu_{\tau}$ signature is given by the decay topology and kinematics of the short lived $\tau^{-}$ leptons produced in the interaction 
of $\nu_{\tau}N\rightarrow\tau^{-}X$ and decaying to one prong ($\mu, e$ or $hadron$) or three prongs, which are~\cite{ref:pdg}:
\begin{eqnarray}
\tau^{-} \rightarrow \mu^{-} \nu_\mu \bar{\nu}_\tau \hspace{2.0cm} with \hspace{0.5cm} BR = 17.36 \pm {0.05}\% \hspace{4.3cm}  \\ 
\tau^{-} \rightarrow e^{-} \nu_e \bar{\nu}_\tau  \hspace{2.1cm} with \hspace{0.5cm} BR = 17.85 \pm {0.05}\% \hspace{4.4cm}  \\
\tau^{-} \rightarrow h^{-} (n\pi^{0}) \bar{\nu}_\tau \hspace{1.45cm} with \hspace{0.5cm} BR = 49.52 \pm {0.07}\% \hspace{4.3cm}  \\
\tau^{-} \rightarrow 2h^{-}h^{+} (n\pi^{0}) \bar{\nu}_\tau \hspace{0.75cm} with \hspace{0.5cm} BR = 15.19 \pm {0.08}\%.\hspace{3.6cm}  
\end{eqnarray}

\subsubsection{First $\nu_\tau$ Candidate:}
The first $\nu_\tau$ candidate was observed in the 2008-2009
data sample and described in detail in~\cite{ref:taupaper}. The
event has seven prongs at the primary vertex, out of
which four are identified as originating from a hadron
and three have a probability lower than 0.1\% of being
 a muon, none being left by an electron. The
parent track exhibits a kink topology and the daughter
track is identified as produced by a hadron through its
interaction. Its impact parameter with respect to the primary
vertex is (55$\pm$4) $\mu$m while the IP is smaller than 7 $\mu$m for the other tracks. Two 
$\gamma$-rays
have been found pointing to the secondary vertex. The
kinematical measurements of the interaction products passed all the 
selection cuts described in detail in the experiment
proposal~\cite{ref:proposal}. 
All the kinematical variables of the event and the cuts applied are given in
Table~\ref{tab:kinematicfirst}. 

The invariant mass of the two observed $\gamma$-rays
is $120 \pm 20 (stat) \pm 35 (syst)$ MeV/c$^2$ supporting the hypothesis that they are
emitted in a $\pi^{0}$ decay. The invariant mass of the charged
decay daughter assumed to be a $\pi^{-}$
and of the two $\gamma$-rays amount to $640^{+125}_{-80} (stat)^{+100}_{-90} (syst)$ MeV/c$^2$, which is compatible with
the $\rho(770)$ mass. So the decay mode of the candidate is consistent with the hypothesis
$\tau^{-}\rightarrow\rho^{-}\nu_{\tau}$.

\begin{table}[htbp!]
  \centering
  \caption{Kinematical variables of the first $\nu_{\tau}$ candidate event with specified
  criteria for $\tau\to$hadron decay channel.}
  \label{tab:kinematicfirst}
  \begin{tabular}{lccc}
    \hline
Variable                &   Measured          & Selection criteria  \\
    \hline
Kink angle (mrad)     & $42\pm 2$           &  \textgreater 20           \\
Angle $\phi$ (deg)    & $173\pm 2$          & \textgreater 90             \\
Decay length ($\mu$m) & $1335\pm 35$        & Within 2 plates     \\
P daughter (GeV/c)    & $12^{+6}_{-3}$      & \textgreater 2             \\
P$_T$ daughter (MeV/c)   & $470^{+230}_{-120}$ & \textgreater 300 ($\gamma$ attached) \\
Missing P$_T$ (MeV/c)    & $570^{+320}_{-170}$ & \textless 1000           \\
    \hline
  \end{tabular}
\end{table}

\begin{figure}[htb]
\begin{center}
\includegraphics[width=0.45\textwidth]{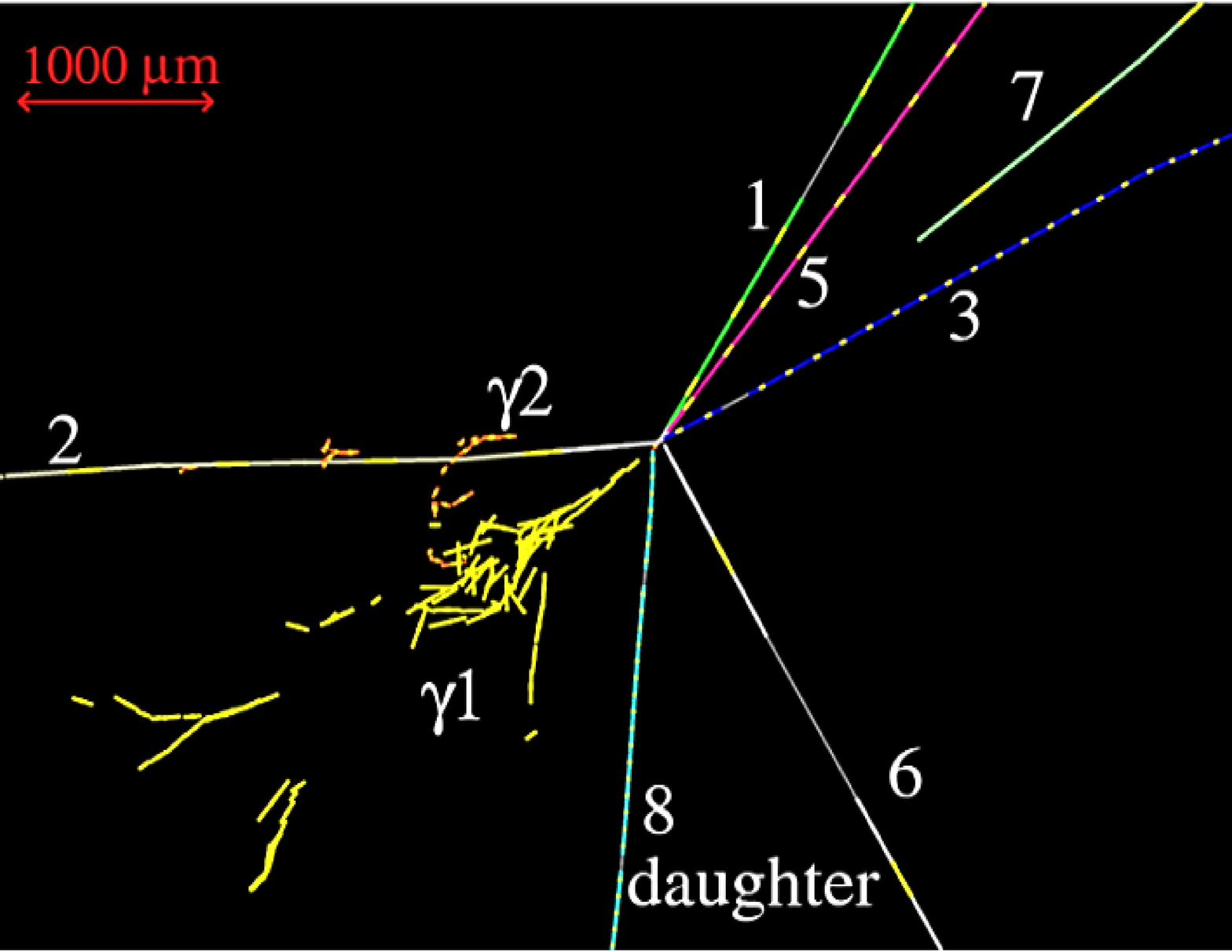}
\hspace{0.5cm}
\includegraphics[width=0.45\textwidth]{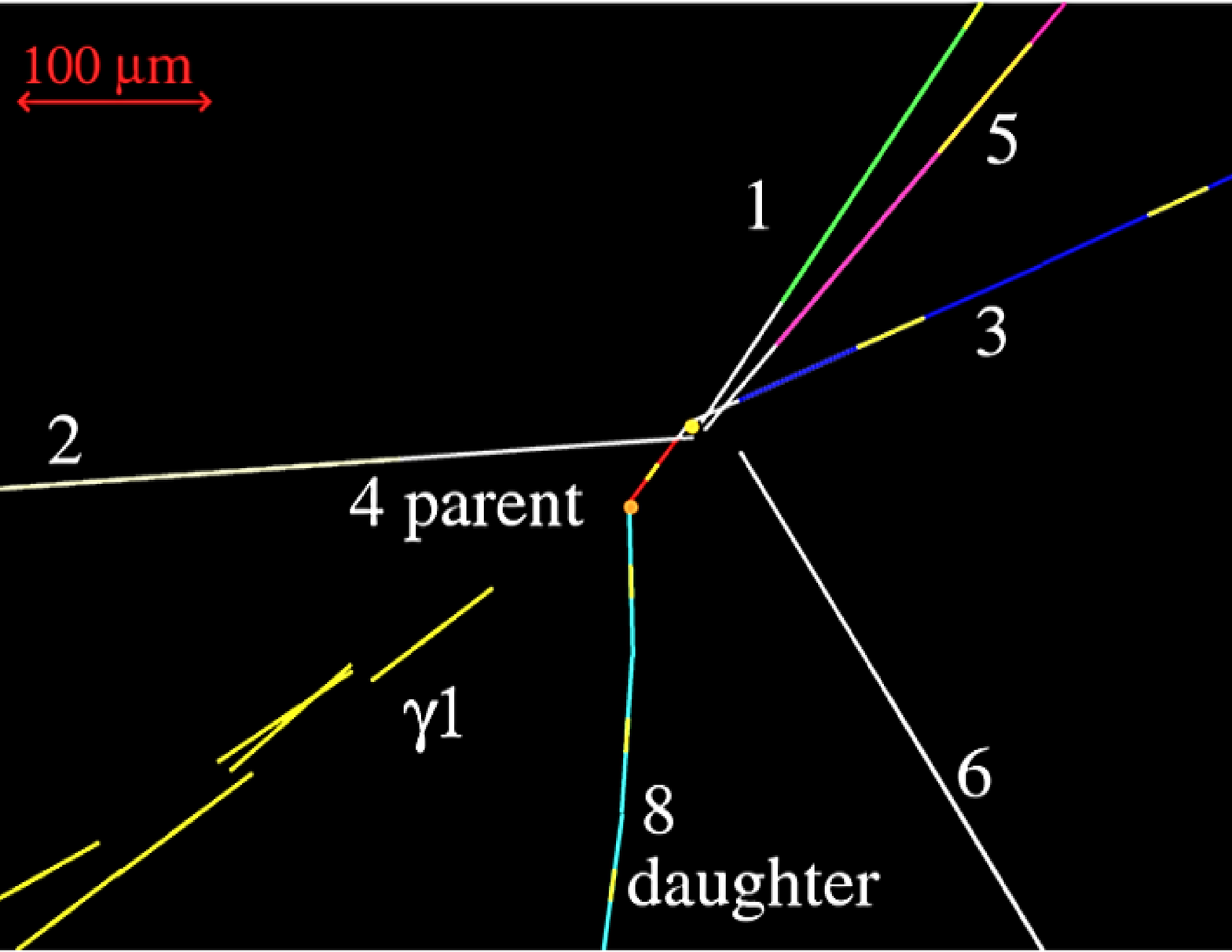}
\end{center}
\caption{Display of the first $\nu_{\tau}$ candidate event. Left: view
transverse to the neutrino direction. Right: same view zoomed on the
vertices.The short track named "4 parent" is the $\tau^{-}$
candidate.}
\label{fig:tau}
\end{figure}

\subsubsection{Second $\nu_\tau$ Candidate:}
\begin{figure}[htb]
\begin{center}
\includegraphics[width=0.45\textwidth]{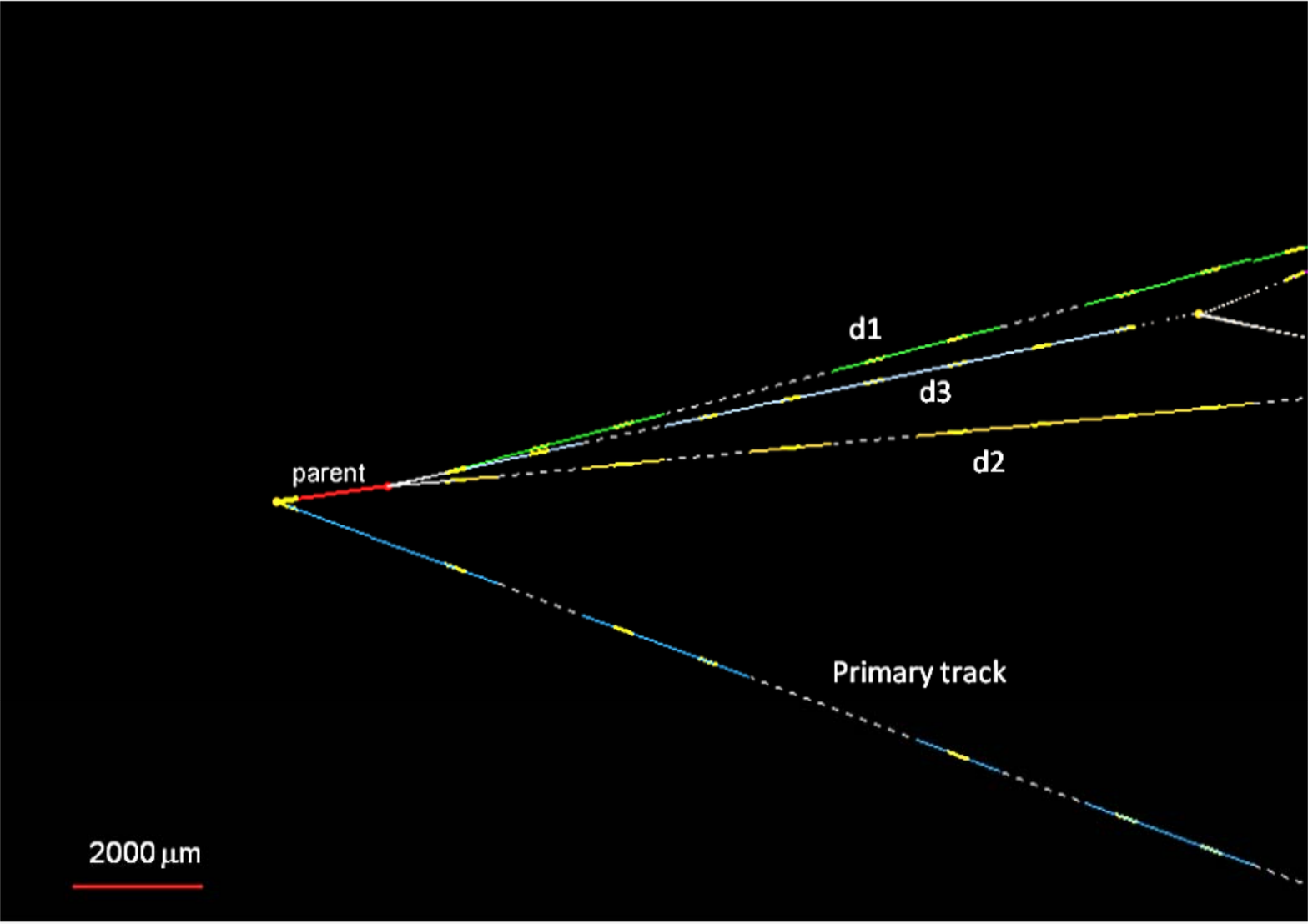}
\hspace{0.5cm}
\includegraphics[width=0.39\textwidth,height=0.24\textheight]{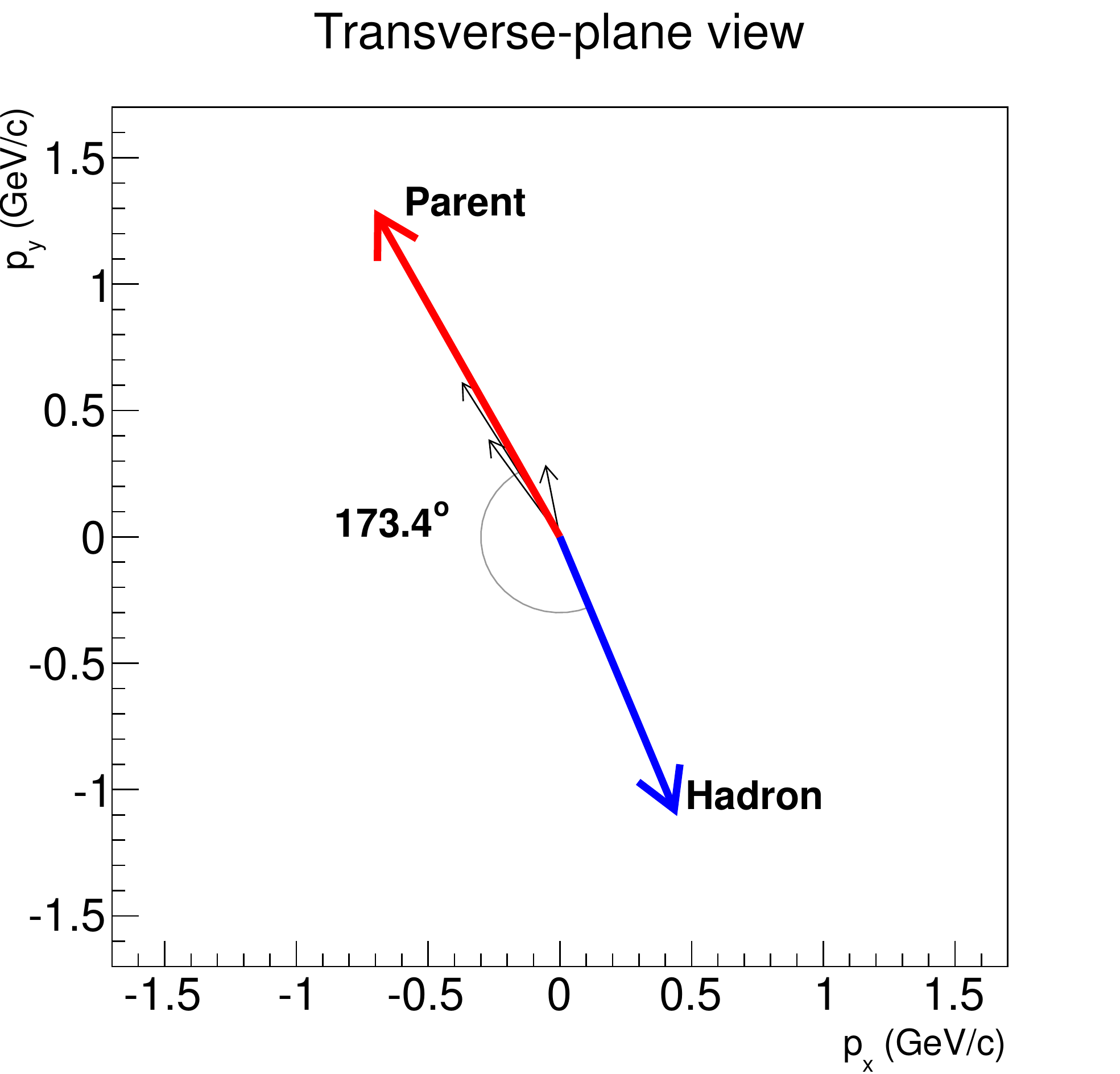}
\end{center}
\caption{Left: Event display of the second $\nu_\tau$ candidate; Right: CNGS transverse plane momentum balancing.}
\label{fig:secondtau}
\end{figure}

The second $\nu_\tau$ candidate was found in the 2011 data
sample. As shown in Figure~\ref{fig:secondtau}, the event is a 2 prong interaction
with production of a short track with flight length
of 1.54 mm (the $\tau$ lepton) and a longer track identified
as a hadron. The interaction vertex lies in the lead plate
between two emulsion films and one nuclear fragment
has been associated to the primary vertex. The decay
vertex of the $\tau$ lepton is in the plastic base in between
the two sensitive layers of an emulsion film. No nuclear fragments have been found pointing to the
decay vertex. The parent track left by the candidate, $\tau$ lepton, decays in three daughter
particles, one of them reinteracting in emulsion after 10 lead plates, $\sim$1.3 cm downstream of the decay point,
and producing 2 charged tracks and
four back-scattered nuclear fragments.
The charged hadron at the primary vertex was followed
downstream and found to stop after
2 walls. The momentum of the track has been evaluated to be
$2.8^{+2.1}_{-2.5}$ GeV/c making highly unfavorable the hypothesis
of a muon on the basis of the consistency between
momentum and range. The daughter particles belonging
to the  decay are also identified as hadrons on the basis
of momentum-range consistency checks. Since there are no muons at the primary vertex,  satisfying all the specified
criteria summarized in Table~\ref{tab:kinematicsecond}, the event is
compatible with a $\tau\to$3 hadrons decay channel candidate.

\begin{table}[htbp!]
\centering
  \caption{Kinematical variables of the second $\nu_{\tau}$ candidate event with specified
  criteria for $\tau\to$3 hadrons decay channel.}
  \label{tab:kinematicsecond}
  \begin{tabular}{lccc}
    \hline
Variable                &   Measured          & Selection criteria  \\
    \hline
Kink angle (mrad)     & $42\pm 2$           &  \textgreater 20           \\
Angle $\phi$ (deg)    & $173\pm 2$          & \textgreater 90             \\
P at 2$^{ndry}$ vertex [GeV/c] & $87.4\pm1.5$ & \textgreater 3.0 \\
P$_{T}$ at 1$^{ry}$ vertex [GeV/c] & $0.31\pm0.11$ & \textless 1.0 \\
Min. invariant mass [GeV/c$^2$] & $0.96\pm0.13$ & 0.5\textless m \textless 2.0 \\
Invariant mass [GeV/c$^2$] & $0.80\pm0.12$ & 0.5\textless m \textless 2.0 \\
    \hline
  \end{tabular}
\end{table}

\subsubsection{Charm Background:}

Charmed particles are produced in about 4\% of $\nu_\mu$ CC interactions at the
CNGS energy. They have lifetimes similar to that of the $\tau$ lepton and have most of the decay
topologies in common. These considerations show how the study of the production of charmed
particles in the OPERA experiment is useful to validate the procedures used for the selection and
identification of $\nu_\tau$ interaction candidates.
The number of charm decays found so far in the data taken from 2008-2011 is 49 with an expectation of 51$\pm$7.5 events. 

The statistical significance of the observation of two 
$\nu_\tau$ candidate events is under evaluation, 
along with the recomputation of efficiencies for signal and background, taking
into account different sample selections. 
The preliminary number of expected $\nu_\tau$ in the analysed
data sample is 2.1, with preliminary background estimation of 0.2 events.

\subsection{$\nu_{\mu}\to\nu_{e}$ oscillations}

Thanks to the capability of electron identification and to the small contamination of $\nu_{e}$
in the CNGS neutrino beam, the experiment has also performed a $\nu_{\mu}\to\nu_{e}$ oscillation search~\cite{ref:nuepaper}.

\begin{figure}[htbp]
\includegraphics[width=0.45\textwidth]{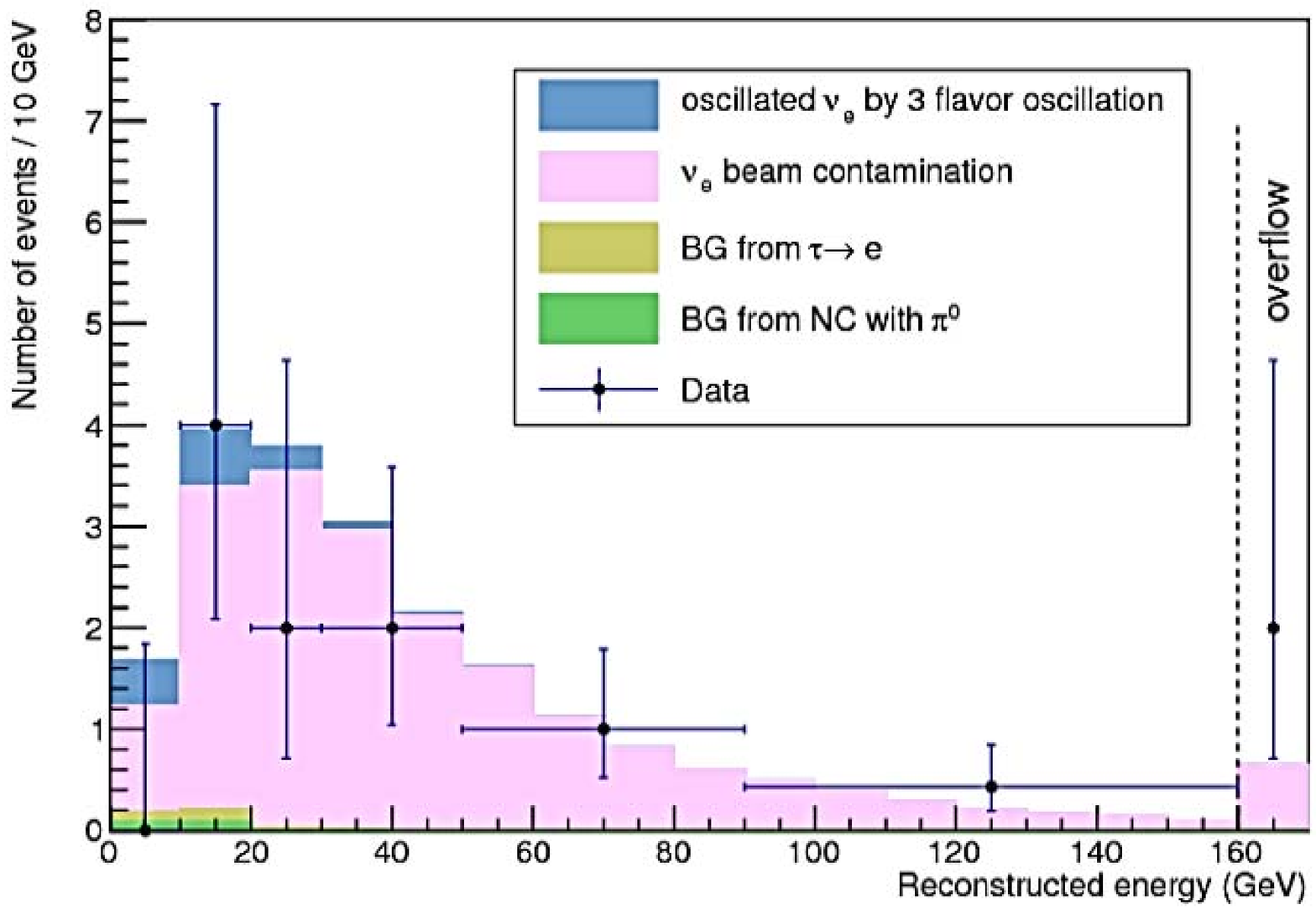}
\hspace{0.5cm}
\includegraphics[width=0.45\textwidth]{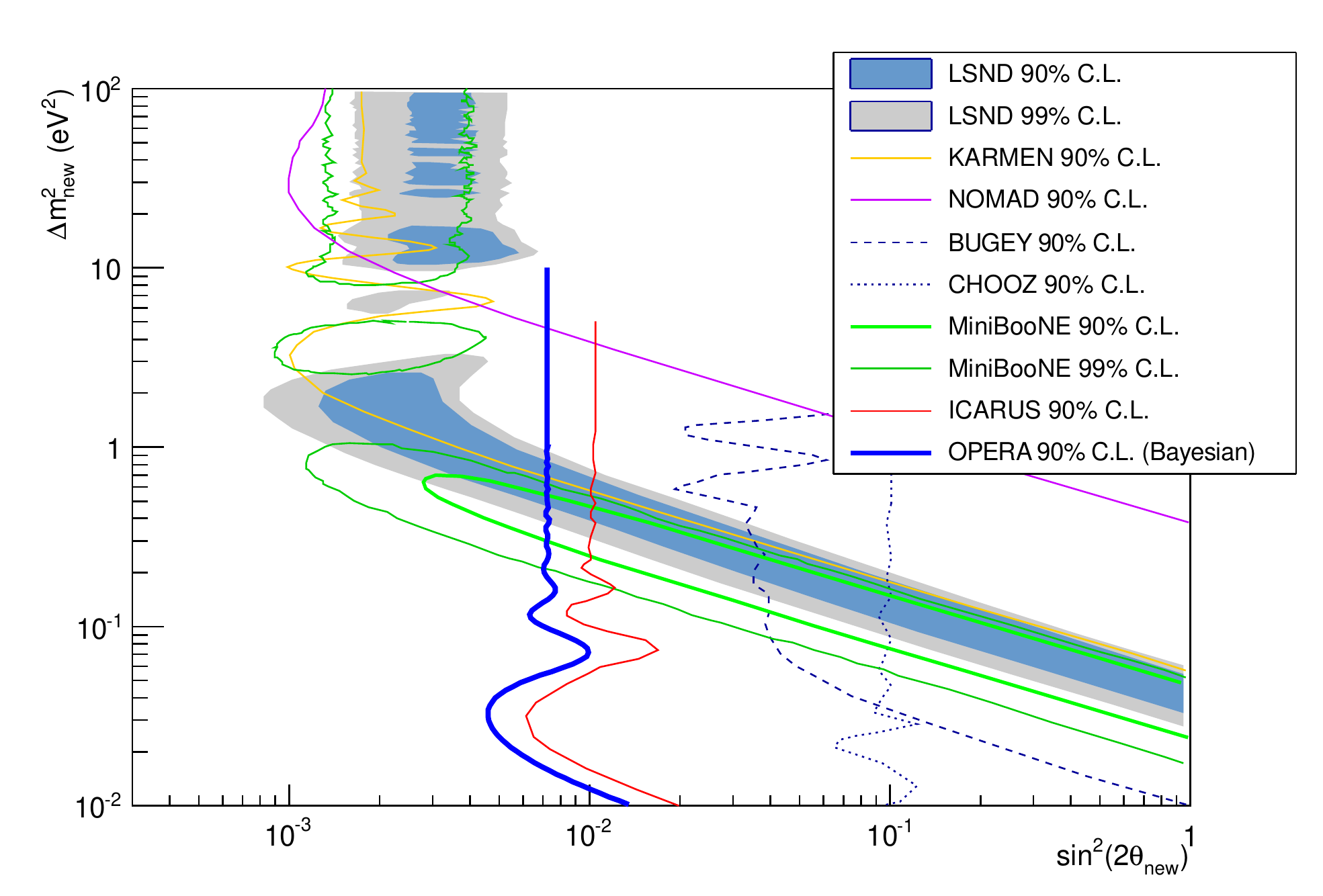}
\caption{Left: Distribution of the reconstructed energy of the $\nu_e$ events and expected spectrum from different
sources, normalized to the number of pots analysed.	
Right:The exclusion plot for the parameters of the non-standard $\nu_\mu\to\nu_e$ oscillation, 
obtained from this analysis using the Bayesian method, is shown. 
The other limits shown, mostly using frequentist methods, are from KARMEN 
($\overline{\nu}_\mu \rightarrow \overline{\nu}_{e}$ \cite{ref:karmen}), 
BUGEY ($\overline{\nu}_{e}$ disappearance \cite{ref:bugey}), 
CHOOZ ($\overline{\nu}_{e}$ disappearance \cite{ref:chooz2}), 
NOMAD ($\nu_{\mu} \rightarrow \nu_{e}$ \cite{ref:nomad}) 
and ICARUS ($\nu_{\mu} \rightarrow \nu_{e}$ \cite{ref:icarus}). 
The regions corresponding to the positive indications reported by  LSND ($\bar{\nu}_{\mu} \rightarrow \bar{\nu}_{e}$ \cite{ref:lsnd}) 
and MiniBooNE  ($\nu_{\mu}\to\nu_{e}$ and $\bar{\nu}_{\mu}\to\bar{\nu}_{e}$ \cite{ref:miniboone}) are also shown.}
\label{fig:exclusion}
\end{figure}

The experiment searched for the appearance of $\nu_e$ 
in the CNGS neutrino beam using the data collected in 2008 and 2009, 
corresponding to an integrated intensity of 5.25 $\times$ 10$^{19}$ pot. 
The systematic $\nu_e$ search was applied on 505 NC-like located events, 
as mentioned in section~\ref{decaysearch}. 19 $\nu_e$ candidate events were
detected with an expectation of 19.4$\pm$2.8 (syst) events from beam contamination and 1.4 event from oscillation.
Apart from the beam contamination, two main sources of background are considered for $\nu_e$ search: (1) 
$\pi^0$ misidentified as an electron in neutrino interactions without a reconstructed muon, and (2) $\nu_\tau$ CC
interactions with the decay of the $\tau$ into an electron.
The first background contribution has been evaluated directly from the data and it was found 
to be 0.2$\pm$0.2 event. While the second one computed by 
MC simulation assuming $\nu_{\mu}\to\nu_{\tau}$ oscillation in a three flavour neutrino mixing scenario
 using up-to-date oscillation parameters; the background was found to be 
0.3$\pm$0.1. 
The total $\nu_e$ background events was estimated to be 19.8$\pm$2.8 (syst). 
This number is in agreement with the 19 observed $\nu_e$ candidate
events. The energy of each event was estimated by applying a correction function (using a simulation 
of $\nu_e$ CC interactions in the OPERA detector to find a correlation between neutrino energy and hadronic energy) on the hadronic energy
measured in the electonic detector.
Figure~\ref{fig:exclusion} shows the reconstructed energy distribution of the 19 $\nu_e$ candidates, compared with
the expected reconstructed energy spectra of the $\nu_e$ beam contamination, the oscillated $\nu_e$ and the backgrounds (1) and (2), normalized
to the number of pots analysed. 

To increase the signal to background ratio, events selection with an energy lower than 20 GeV was applied. 
Within this cut,
we observed 4 events with an expectation of 4.2 events from $\nu_e$ beam contamination and 0.4 events 
from backgrounds (1) and (2). The number of 
observed events is compatible with the non-oscillation hypothesis and yields an upper limit $sin^{2}2\theta_{13} < $ 0.44 (90\% C.L.).

OPERA sets an upper limit for a non-standard $\nu_e$ appearance in the parameter space suggested 
by the results of the LSND~\cite{ref:lsnd} and MiniBooNE~\cite{ref:miniboone} experiments. 
One mass scale dominance approximation with new oscillation
parameters $\theta_{new}$ and large $\Delta{m}^2_{new}$ ($>$0.1\eV2) have been considered.
An expectation of 9.4$\pm$1.3 $(syst)$ events was made for an
energy $<$ 30 GeV (optimal cut on the reconstructed energy in terms of sensitivity), 
while 6 events have been found in the data.

The 90\% C.L. upper limit on $sin^{2}2\theta_{new}$ is computed by comparing
 the expectation from oscillation and background, with the observed number of events. 

Given the underfluctuation of the data, a Bayesian statistical approach has been 
 used to determine the upper limit and the exclusion plot shown in Figure~\ref{fig:exclusion}.
For comparison, results from other experiments, working at different $L/E$ regimes, are
also reported in this figure.
 For large $\Delta m^{2}_{new}$ values, the 90\% C.L. upper 
 limit on $sin^{2}2\theta_{new}$ reaches $7.2 \times 10^{-3}$, while the sensitivity corresponding to 
the analysed statistics is $10.4 \times 10^{-3}$.

\section{Conclusions}
The physics run of the OPERA experiment started in 2008 and ended on December 2012. In total
17.97$\times10^{19}$ pots were collected, corresponding to about 80\% forseen value of $22.5\times10^{19}$ pots.
So far 4505 neutrino interactions have been fully analysed looking for tau lepton 
decay topologies and 3 $\nu_\tau$ candidate events have been observed.

A systematic search for $\nu_e$ CC interactions was performed searching for sub leading 
$\nu_{\mu}\to\nu_e$ oscillations. Using $5.3\times10^{19}$ pots with 505 located NC-like events from 2008-2009 data, 19 $\nu_e$
candidates have been detected with an expectation of 19.8$\pm$2.8 (syst) events, a result  compatible 
with the non-oscillation hypothesis. Using the same data sample, OPERA set an upper limit in the parameter space available
for a non-standard $\nu_e$ appearance indicated by LSND and MiniBooNE experiments.

\title
\end{document}